\documentclass[12pt]{spieman}

\usepackage[cmex10]{amsmath}
\usepackage{graphicx}  
\usepackage{hyperref}  
\usepackage{setspace}
\usepackage{tocloft}
\usepackage{subcaption}
\usepackage{listings}
\newcommand{\srccode}[1]{\textit{#1}}
\newcommand{\ket}[1]{| #1 \rangle}

\title{Software-defined Quantum Communication Systems}

\author{Travis S. Humble\supscr{a,b} and Ronald J. Sadlier\supscr{a,c}}

\affiliation{\supscrsm{a} Quantum Computing Institute, Oak Ridge National Laboratory, Oak Ridge, TN, USA \\
\supscrsm{b} Bredesen Center, University of Tennessee, Knoxville, TN, USA\\
\supscrsm{c} University of Rhode Island, Kingston, RI, USA
}

\cftpagenumbersoff{figure}

\begin{document}

\maketitle

\begin{abstract}
Quantum communication systems harness modern physics through state-of-the-art optical engineering to provide revolutionary capabilities. An important concern for quantum communication engineering is designing and prototyping these systems to evaluate proposed capabilities. We apply the paradigm of software-defined communication for engineering quantum communication systems to facilitate rapid prototyping and prototype comparisons. We detail how to decompose quantum communication terminals into functional layers defining hardware, software, and middleware concerns, and we describe how each layer behaves. Using the super-dense coding protocol as an example, we describe implementations of both the transmitter and receiver, and we present results from numerical simulations of the behavior. We conclude that software-defined quantum communication provides a robust framework in which to explore the large design space offered by this new regime of communication.
\end{abstract}


{\noindent \footnotesize{\bf Address all correspondence to}: Travis Humble, Oak Ridge National Laboratory, MS 6015, One Bethel Valley Road, Oak Ridge, TN, 37831-6015  USA; Tel: +1 865-574-6162; E-mail:  \linkable{humblets@ornl.gov} }


\begin{spacing}{2}   

\section{Introduction}
Quantum communication (QC) is an active area of fundamental research and technology development that makes use of the quantum properties of light to transmit and receive quantum information. \cite{QCOMM2009,COMSOC2013} It enables novel capabilities like quantum teleportation or quantum key distribution that cannot be provided by means of classical communication \cite{Bennett1993, BB84}. The design of prototype QC systems is an important step towards realizing theoretical predictions and assessing experimental performance. Of course, similar issues face classical communication (CC) systems and we may expect QC research to leverage existing methods for system prototyping. In particular, software-defined implementations have proven useful for providing flexibility in the design and testing of conventional radio systems.\cite{Mitola2000} In this contribution, we extend the software-defined communication paradigm to the design and development of QC systems.\cite{HumbleSadlier2013}
\par
Software-defined communication (SDC) allocates signal processing tasks that nominally require specialized hardware to software implementations based on general-purpose computational power.\cite{Mitola2000} For example, within traditional radio communications, the ideal SDC receiver would use an antenna and analog-to-digital converter (ADC) for signal sampling before handing off the remaining waveform processing tasks to software. These tasks, including mixing, filtering, and demodulation, are then tuned by simply reprogramming the radio. Reprogrammable radios promise to be cheaper to design and build than using dedicated and fixed hardware components. More important, the ability for SDC to configure itself in real time affords the opportunity to adapt to the transmission environment, i.e., a cognitive radio.\cite{Mitola1999} The SDC paradigm is not restricted to radios; similar ideas have been argued for use in optical communication systems. \cite{Savory2008, Cox2011}
\par
Although much of the physics underlying quantum communication is very different from conventional communications, the SDC paradigm can apply to building QC systems as well. This is because both domains employ many of the same processing primitives at the information (bit) level. This includes the de/modulation and de/coding techniques required for individual transmissions in addition to the handshaking exchanges needed to negotiate complete protocols. These common needs motivate our consideration of software-defined quantum communication systems and our evaluation of its feasibility with state of the art quantum optical hardware. 
\par
Of course, there are notable differences between quantum communication (QC) and classical communication (CC). These differences manifest from how information is encoded into the photonic carrier. In particular, QC encodes information into the quantum state of a photon using any number of degrees of freedom, e.g., polarization, quadrature phase, spatial mode, angular momentum, frequency, etc. By comparison, CC uses macroscopic amounts of photons to encode the classical state of the sames degrees of freedom. This difference leads to unique capabilities for each physical domain. \cite{Wilde2011}
\par
Notwithstanding differences at the physical layer, quantum and classical communication share a dependence on logical control data known as {\it metadata}. Both regimes require metadata to control, manage, organize and annotate the transmitted payload. In a typical CC example, metadata may be concatenated with the payload by the transmitter and then extracted by the receiver.  This information may, for example, identify the demodulation needed to recover the payload or specify the destination address needed for routing.
\par
In the case of QC, classical metadata may either be shared through a synchronized side-channel or generated by measurement of the transmitted quantum state. An example of the latter is found in quantum key distribution (QKD) in which the transmitter and receiver share measurement results to determine the next steps in the key generation protocol.\cite{BB84} In the QKD example, some measurements serve the role of metadata while others represent the payload. These distinctions are not known at the time of transmission but are derived using an agreed upon CC protocol. By contrast, quantum teleportation and entanglement swapping typically require a side channel through which to share the classical measurements recorded by the transmitter and needed by the receiver to recover or relay the quantum state.\cite{Bennett1993} Similar examples include the cases of quantum memory modules or quantum routers that use dynamic addresses to store \cite{Simon2010} and route information \cite{Herbauts2013}, respectively. Moreover, Fujiwara has shown how metadata may even be encoded into the quantum state, which would move our software paradigm into a quantum computational setting. \cite{Fujiwara2013}. These latter examples serve to emphasize that a quantum receiver need only operate on the transmitted states and not necessarily measure them. It is also possible to process metadata within the quantum receiver hardware. This approach has been taken previously in some QKD and quantum teleportation testbeds. \cite{Lodewyck2007, Jouguet2012, Zhang2012, Jesus2012, Cui2013}
\par
The ubiquity and importance of metadata in QC motivates consideration of how the SDC paradigm may be leveraged to build  prototype systems. We will show that a typical QC transceiver can be decomposed into components that separate the physical encoding layer from the metadata control layer. These layers can then be identified as separating the concerns between the hardware and software domains while a third middleware layer interpolates between these domains. We describe implementations of all three domains that maintain a natural separation of concerns while also providing a tunable interface for QC developers. 
\par
In this paper, we present a framework for defining a software-defined QC system with respect to hardware, middleware, and software layers. We elaborate on the abstraction of these different layers and provide a concrete example for the case of a point-to-point super-dense coding communication system. We include details of how the complete system can be constructed and emphasize how the software and middleware layers should interact in order to make the physics oblivious to an end user.
\section{Framework}
\label{sec:frame}
\begin{figure}[h]
\begin{center}
\includegraphics[scale=.4]{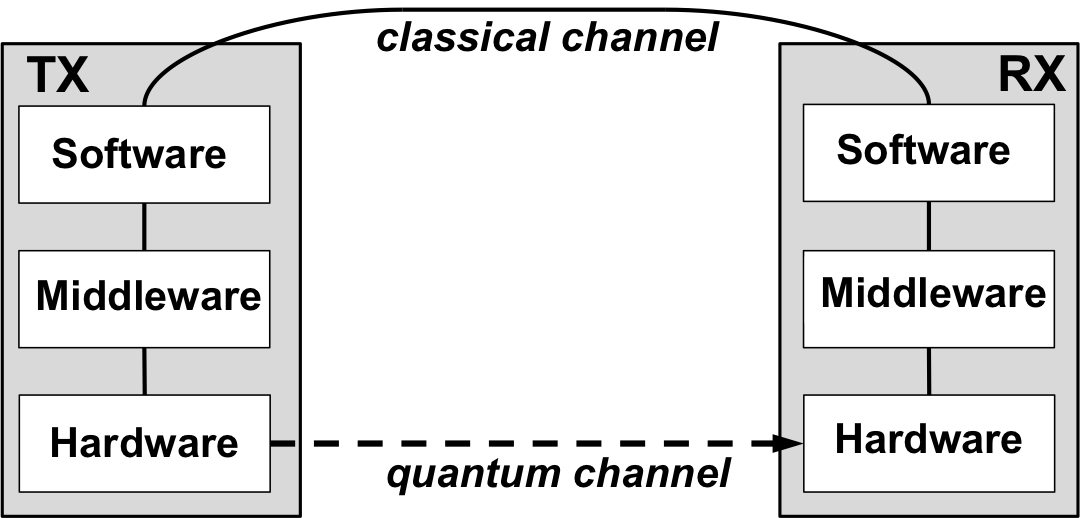}
\caption{Decomposition of an SDQC system consisting of a single transmitter-receiver pair.  Each terminal is composed from hardware, software, and middleware layers. Hardware layers interact via a quantum channel while software layers interact over the classical channel. The middleware serves to translate between the languages serving the hardware and software domains.}
\label{SDQC_LAYERS}
\end{center}
\end{figure}

We formalize the software-defined quantum communication (SDQC) framework by considering a single transmitter-receiver pair with a quantum transmitter (TX) and quantum receiver (RX). A decomposition of each terminal is shown in Fig. \ref{SDQC_LAYERS} with respect to the functional domain layers.  These layers serve to separate development concerns in constructing each transceiver with respect to the hardware physics, the software protocol, and a middleware that communicates between the two domains. Similar decompositions can be applied to previously developed QC systems. Our objective is to show how to deliberately identify these domains at an abstract level and subsequently develop them into concrete realizations. 
%
\subsection{Transmitter and Receiver Structure}
\par
\begin{figure}[h]
\begin{center}
\includegraphics[scale=.4]{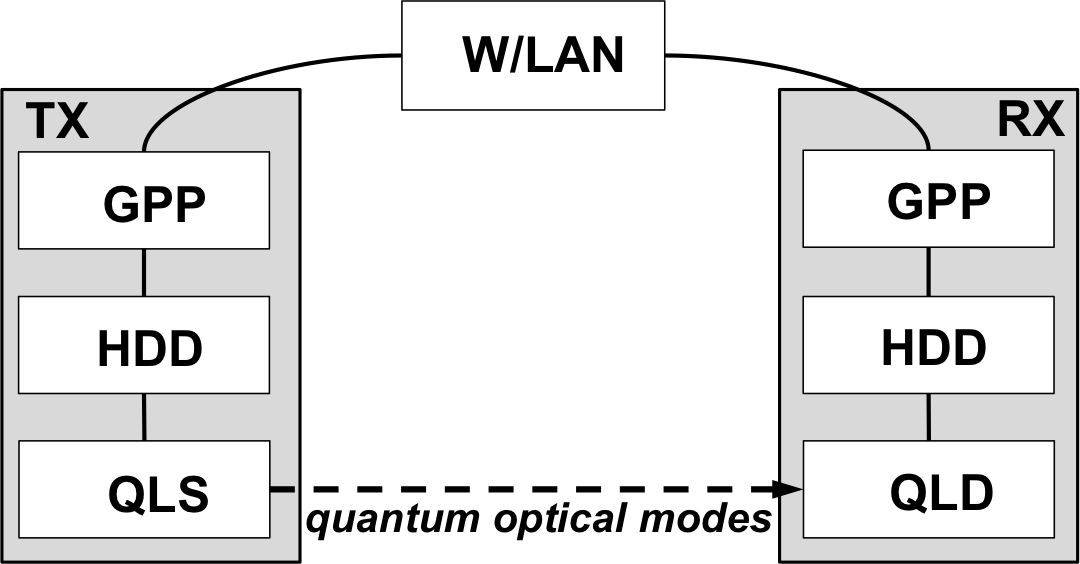}
\caption{A component representation of the SDQC system shown in Fig.~\ref{SDQC_LAYERS}: a transmitter (TX) consists of a quantum light source (QLS) driven by a hardware device driver (HDD) that is controlled by a general purpose processor (GPP). The TX GPP communicates over a wide/local area network (W/LAN) with a receiver RX. The RX GPP manages an HDD that monitors a quantum light detector (QLD). The QLS/QLD link defines the quantum channel.}
\label{SDQC_ARCH}
\end{center}
\end{figure}

A concrete representation of the SDQC framework is shown in Fig.~\ref{SDQC_ARCH}, in which the TX hardware layer is expressed as a quantum light source (QLS) for preparing quantum states and accessing the quantum channel, the middleware is represented as a hardware device driver (HDD), and the software layer is represented by a general purpose processor (GPP) running a user-defined QC program. The classical channel is assumed to be a local area network (LAN) while the quantum channel is represented by some quantum optical modes.
\par
In the TX of Fig.~\ref{SDQC_ARCH}, the prepared states are encoded into the Hilbert (sub)space of some photonic degrees of freedom. Candidates for the encoding include the polarization, orbital angular momentum, or field quadrature variables among others. The hardware layer is modeled to include all components necessary for state preparation such as polarization or phase modulators, with the physical encoding controlled by the HDD. It is the middleware that implements the interface to the QLS for use by software. The software issues controls and manages the TX behavior by signaling to the QLS which states to prepare.  As a simple example, software can send a bit to the HDD specifying the basis to use for state preparation. The HDD middleware may then parse this bit into the appropriate sequence of QLS control signals. Of course, more elaborate protocols will require more elaborate interactions between the two layers, but in general the middleware and hardware do not require detailed information about the protocol implemented in software.
\par
The RX in Fig.~\ref{SDQC_ARCH} is modeled similar to the TX, except that the RX software now drives a quantum light detector (QLD). The QLD measures received photons and outputs measurement information. The RX middleware serves to sample the measurement information and relay it back to the software. It is the presence of the QLD which distinguishes the RX from TX. A transceiver (TRX) combining both QLS and QLD components would need only one middleware interface to implement this design. 
\par
For both the TX and RX, the software layer also serves to communicate required metadata over the LAN. This includes, for example, negotiating the key protocol inherent to QKD or relaying feed-forward measurement information for quantum teleportation. Because the software is assumed to be reprogrammable, techniques used in sharing metadata can be modified by the end users as needed. As an example, classical error correction steps are important to deriving keys in QKD but the error codes used may require tuning to match the channel and observed bit error rates. \cite{Johnson2013} These types of modifications are easily made using software-defined implementations of the user's selected protocol.
\subsection{Hardware Layer}
The hardware layer expresses components that are fundamental to the physical encoding of quantum information into the transmitted signal.
Many quantum light sources and detectors are available as off-the-shelf components. For example, single-photon detectors are sufficiently advanced and wide-spread in their application as to be stand-alone items from optical suppliers. Similarly, weak-coherent pulses generated from attenuated output of photodiodes are easily setup for transmission. There is a significant variety in these elements with respect to wavelength, bandwidth, stability, and cost so as to warrant their consideration as a replaceable element in the QLS/D design. Individual applications require suitable pairing between the wavelengths of the source and detector but the modularity of system design ensures such hardware changes do not undermine the software and middleware layer.  Similar arguments also hold for research-grade hardware that may be tailored for specific experimental questions. The essential similarity is that both require externally accessible interfaces for the actively controlled elements and generated metadata.
\par
The engineering challenging to the development of the QLS/D hardware within the SDQC framework is correctly executing the controls sent to the hardware layer. Nominally, the SDQC design implies that the hardware consists of programmable elements that may be driven explicitly by the middleware. Device drivers supplied with most actively controlled components, e.g., translation stages, piezo-electric controllers, phase modulators etc., satisfy this requirement. Collectively, these device drivers and control wires define the hardware interface. The remaining challenge, therefore, is the integration and mapping of hardware control implementations into a well defined interface. For most lab-based QC experiments, this is traditionally  accomplished in an ad hoc manner that is sufficient for proof of principle but not robust to updates or modifications. Within SDQC, it is the role of the middleware to ease the hardware management by abstracting the interface required by software layer while enforcing the constraints imposed by hardware specification.
\subsection{Middleware Layer}
The middleware parses metadata within the TX and RX. This includes translating metadata generated by the TX software specifying which qubits (states) to prepare within the hardware as well as tagging raw measurement data generated by the RX hardware. A middleware interface is defined to separate the concerns between the structure of the hardware and its expected behaviors required by the software.
\par
Implementing the middleware requires knowledge of what hardware components are available and the means by which they are controlled, e.g., via specified device drivers. Several controlled components may be synthesized to implement selected software behavior, for example, state preparation or measurement in a specific basis. However, the particular methods implemented by the middleware to manage control of the hardware should be hidden from software in order to maintain separation of concerns. In addition, the middleware need only provide a library of elementary functions that can be called upon. This separates the middleware from whatever particular protocol is being implemented. Similarly, the middleware relays information up to the software but it remains oblivious to its usage. 
\par
The promotion of metadata from the hardware layer to the software layer requires translation between the domain specific languages native to those layers. This is built into the design of the middleware interface and determined by the level of abstraction provided. The middleware interface can and should vary with the intended use cases of the terminal. For example, a terminal could be designed such that a user-developed software is able to explicitly request that the middleware ``rotate waveplate 1 to angle $\theta = \pi/4$". The resulting middleware implementation would then relay the appropriately parsed signal to the hardware in order to prepare the specified configuration. Alternatively, the middleware may be designed to accept only more abstract commands, e.g., ``prepare a qubit in the $X$ basis", in which case the translation into the hardware language would be determined by a more sophisticated middleware implementation that included rotation of the necessary waveplates. These cases are distinguished by how much they abstract away the hardware components from the software protocols. Either approach may be a useful implementation - the best choice is driven by the expected needs of the end user.
\subsection{Software Layer}
In the SDQC framework, the software layer defines the abstracted behavior of the hardware but not the implementation details. The level of abstraction and therefore control that is provided to the software layer is determined by the overall design of the terminal and especially the limitations implied by the middleware interface. Depending on these design decisions, the software layer may explicitly define the type of information to be communicated as well as methods for validating transmission and negotiating classical metadata between the TX and RX. Alternatively, the middleware interface may only provide access to a more limited set of behaviors, for example, how many bits to exchange between users. The flexibility in assigning these responsibilities offers a natural way to control the terminal design space.
\par
It seems necessary to justify that the demands of existing and near-term prototype QC systems can be satisfied using software control. Current state of the art QC systems provide at most detection at rates of 1 Gbit/sec \cite{Eisaman2011}. This upper bound on bit rate is due largely to operational limits of current light sources, which must employ trade-offs between quantum detector efficiency and response time. Additional losses arising from transmission only serve to reduce observed count rates and further limit QC systems to sub-GHz rates.  By comparison, modern processors containing multiple cores have theoretical clock rates well above 10 GHz. This represents a more than 10-fold increase in processing speed over data acquisition rates. Moreover, these clock rates correspond with $10^{9}$ floating-point operations per second (1 GFLOPS) even for commodity processors. Alongside gigabit per second (Gbps) communication links, the availability of more than 1 GFLOPS suggest it is both possible and reasonable to carry out the computationally intensive part of many QC protocols relatively easily in near real time. Of course, for detection rates beyond 1 Gbps, using off-the-shelf processors may require additional design considerations. For example, the inclusion of specialized coprocessors such as graphical processing units (GPU's) or field-programmable array's (FPGA's) remains an option. However, for the purpose of building reprogrammable  QC systems capable of testing new protocols, it seems modern processor  technology is well matched for prototyping.
\par
The design of the software layer requires a clear specification of the abstraction intended for the application programming infrastructure. This includes the application programming interface (API) exposed to the user as well as the supporting libraries providing the interface with the middleware. This can be accomplished using standard system software programming and device drivers as well as more elaborate integrated programming environments.
\section{Super-Dense Coding System Design and Implementation}
As a demonstration of the SDQC framework, we present an implementation of super-dense coding \cite{Bennett1992}. Super-dense coding is a protocol whereby two users, Alice and Bob, begin by sharing a pair of entangled two-level systems, i.e., qubits. The entangled qubits are initially prepared in the  state
\begin{equation}
\ket{\Phi^{(+)}} = \frac{1}{\sqrt{2}}\left(\ket{0_A, 0_B} + \ket{1_A, 1_B} \right),
\end{equation}
where subscript $A$ denotes Alice's qubit and $B$ denotes Bob's qubit. Alice has an $2$-bit message $b_1b_2$ which she transmits to Bob by applying to her qubit one of the four unitary operators $\mathcal{O} \in \{I, X, Z, XZ\}$. These operators have the distinction of mapping the original state within the complete set of Bell states,
\begin{equation}
\label{eq:bell}
\arraycolsep=1.4pt\def\arraystretch{2.2}
\begin{array}{c}
\ket{\Phi^{(\pm)} } = \frac{1}{\sqrt{2}} \left(\ket{0_A, 0_B} \pm \ket{1_A, 1_B}\right) \\
\ket{\Psi^{(\pm)} } = \frac{1}{\sqrt{2}} \left(\ket{0_A, 1_B} \pm \ket{1_A, 0_B}\right) \\
\end{array}
\end{equation}
The mapping between operators and bit pairs is established by Alice and Bob before beginning the protocol. We will use the mapping:
\begin{equation}
\label{eq:sdcmap}
\begin{tabular}{|c|c|c|}
\hline
$b_1 b_2$ & $\mathcal{O}$ & $\ket{\psi_{A,B}}$ \\ \hline
00 &  I    & $\ket{\Phi^{(+)}}$ \\ \hline 
01 & X    & $\ket{\Psi^{(+)}}$ \\ \hline 
10 & Z    & $\ket{\Phi^{(-)}}$ \\ \hline 
11 & XZ  & $\ket{\Psi^{(-)}}$ \\ \hline 
\end{tabular}
\end{equation}
where $\ket{\psi_{A,B}}$ denotes the state prepared by Alice. After applying the operator $\mathcal{O}$ to her qubit, Alice transmits her qubit to Bob. Upon receiving Alice's qubit, Bob performs a joint measurement that discriminates between the four Bell states. Based on the outcome of the measurement, Bob decodes the original two bits of message.
\subsection{Software Layer}
For our implementation of super-dense coding,  the software layer is a library built within the GNU Radio signal processing framework. GNU Radio is a free software toolkit for deploying software-defined communications systems that offers primitive signal processing blocks for application development \cite{GNR}. We have leveraged GNU Radio by creating the Quantum Information Tool Kit for Application Testing (QITKAT), a library extension that provides both C++ and Python based processing blocks to support prototyping stream-based quantum communication. The QITKAT library includes primitives for expressing communication protocols completely in software. This includes methods for encoding and decoding the SDC messages as well as interfaces exchanging network metadata between users. These blocks can then be connected using an inter-process communication system provided by the GNU Radio run-time environment. The run-time manager is responsible for maintaining the flow of data, while the block developer is responsible for ensuring each blocks consumes and processes samples in the desired way.
\par
\begin{figure}[h]
\begin{center}
\includegraphics[width=0.9\textwidth]{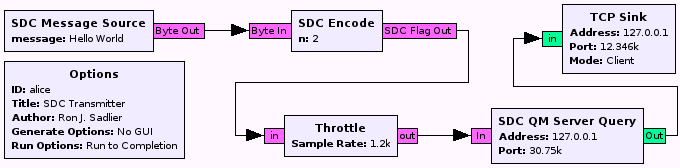}
\caption{A QITKAT flow graph showing the super-dense coding transmitter.}
\label{fig:alice}
\end{center}
\end{figure}
\begin{figure}[h]
\begin{center}
\includegraphics[width=0.9\textwidth]{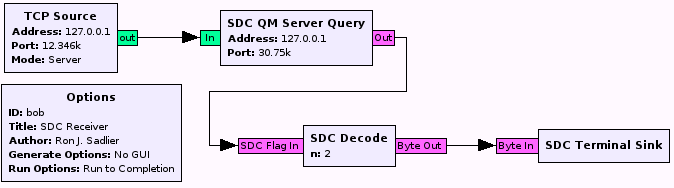}
\caption{A QITKAT flow graph showing the super-dense coding receiver.}
\label{fig:bob}
\end{center}
\end{figure}

Using QITKAT and GNU Radio blocks, we have developed TX and RX programs that permit Alice to encode binary data and send modulated entangled states to Bob, who decodes these modulations from measurements made on the entangled state. The processing flow graphs for the dense coding system are shown in Figs.~\ref{fig:alice} and \ref{fig:bob}. These diagrams describe the flow of information between the processing blocks within the system. In particular, the \srccode{SDC Message Source} block forwards binary strings to the \srccode{SDC Encode} block, which looks up the appropriate operator based on the bit values according to the table in Eq.~(\ref{eq:sdcmap}). The corresponding operator flag is sent to the \srccode{QM Server} block, which represents the middleware interface responsible for translating operators into correct actions on the fiducial Bell state. These commands are issued over a network using TCP messaging.

\par
On the RX end, the \srccode{QM Server} block accepts messages returned by the hardware layer via the same TCP messaging. These 'messages' from the hardware are actually signatures corresponding to specific detection events. After interpreting these events as specific state labels, the \srccode{SDC Decode} block decodes these Bell-state measurement into bit pairs. These bits are then forwarded into the \srccode{SDC Message Sink} block, which serves the purpose of buffering the complete message.
\subsection{Middleware Layer}
The \srccode{QM Server} block serves as a visible middleware component. The encode and decode blocks issue control commands to modulate and measure the Bell state, respectively. The modulations are based on application of the operator $\mathcal{O}$ in Eq.~(\ref{eq:sdcmap}) while the measurements correspond with projections in the Bell basis of Eq.~(\ref{eq:bell}). This block is also responsible for the handshaking between the encode and decode blocks, which in our implementation is simply a classical transmission of packet counter to monitor the qubit sequence. This is in addition to the handshaking that underlies the classical network communications. In the current implementation, the server resides on a separate computer and communication is managed using TCP packets. The \srccode{QM server} may be running local on the same host as either TX or RX clients, or on a separate device as would be a more natural case when the server is managing separate hardware. 
\subsection{Hardware Layer}
For SDC, the necessary hardware includes a source of entangled particles, a modulation mechanism, and a measurement apparatus. We will assume the use of polarization-entangled photon pair states, in which the horizontal and vertical polarizations of the photons are used to encode $ket{0}$ and $\ket{1}$, respectively. A non-deterministic source of polarization-entangled photon pairs can be constructed using the process of spontaneous parametric down conversion (SPDC) pumped by an external laser. This approach, however, lacks a means of announcing the photon's presence. Heralded pair production offers a slightly more complicated alternative but it has the advantage that each photon is tagged as being in a known time slot.\cite{Barz2010}
\par
For polarization entangled biphoton states, the modulation operators are implemented using an optical wave plate for the $X$, $Z$, and $XZ$ transformations. Because the orientation determines the operator being implemented, we can mount the waveplate(s) on an electronically driven rotator \cite{Shelton2011,Rakonjac2013}. The state of the rotator, and the photon polarization, can then be driven using computer-controlled electrical signals. The measurement of the photon pair state at the RX can be implemented partially using linear-optical Bell-state measurement device \cite{Dusek2001}. In this setup, a static beam splitter interferes the two photons and polarization analyzers measurement the resulting state. The observed measurements can then identify 3 of the 4 possible Bell states, but cannot detect all of them.\cite{LCS1999} Alternative approaches to measure all four states come at the cost of additional complexity. \cite{Zaidi2013} In our design, we assume a static optical network precedes a bank of detectors, which output a unique signature for each encoded state.
\subsection{Integration}
We have realized our design using in an FPGA board with an embedded ARM processor. Our implementation is based on a Xilinx Zynq system-on-a-chip board appended with a custom daughter board that collects TTL inputs from a bank of Si-APD detectors. The Zynq board supports communication over Ethernet as well as serial lines or other inputs, and it can be programmed using Xilinx's build tools. The Zynq board serves as a convenient platform for integrating together the FPGA-based control and parsing of electrical signals with the reprogrammable behavior of the ARM processor.
\par
\begin{figure}[h]
\begin{center}
\includegraphics[width=.7\textwidth]{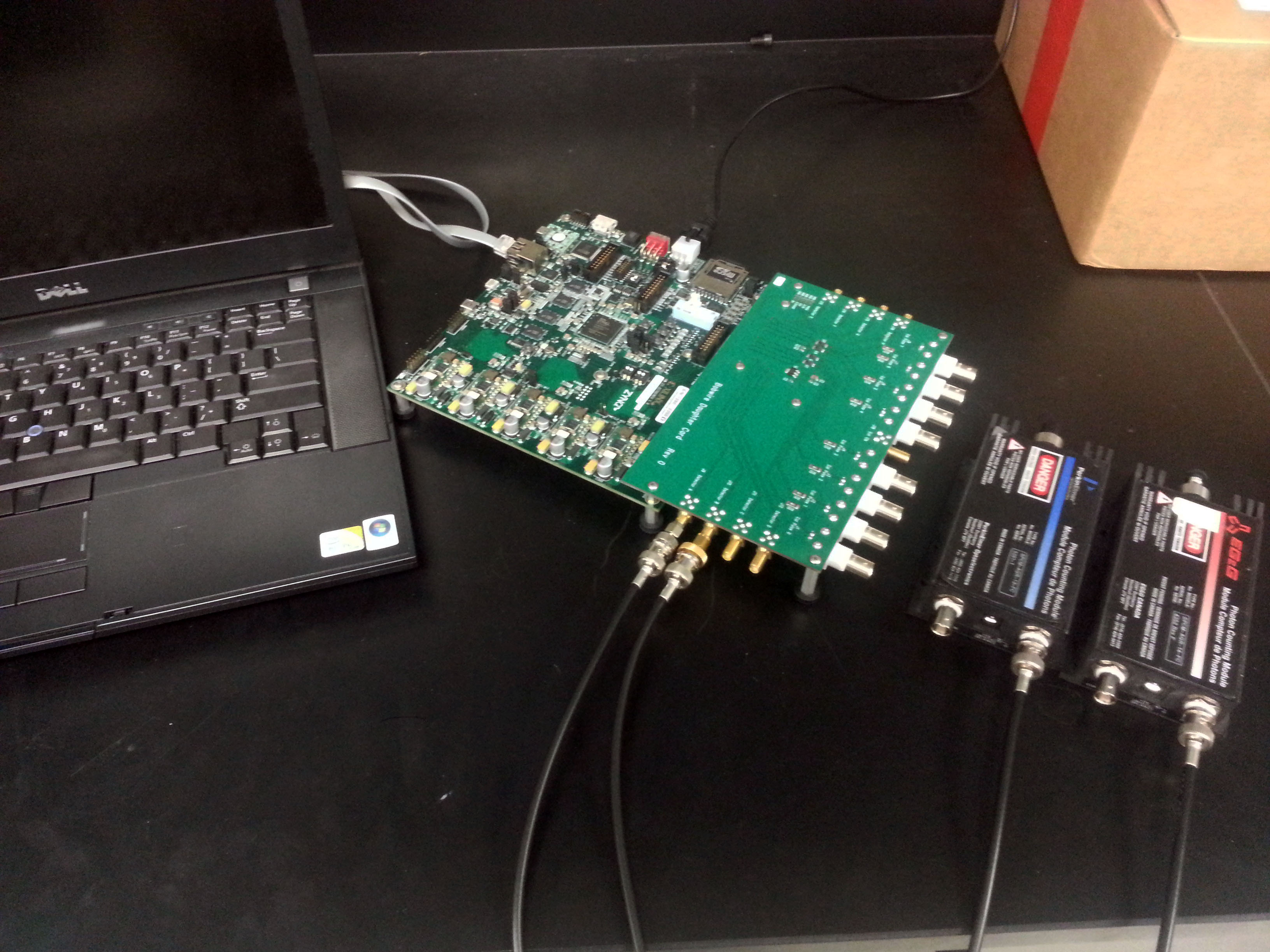}
\caption{A physical representation of the SDQC architecture for the SDC RX implementation. The computer on the left runs the QITKAT program while the customized Zynq board in the middle represents the middleware implementation. On the right, a pair of silicon photodetectors represent the RX hardware and connect to the daughter board.}
\label{fig:sdchw}
\end{center}
\end{figure}

An example of the complete implementation for the RX is shown in Fig.~\ref{fig:sdchw}. In this design, the FPGA accepts TTL signals from detectors connected through the custom daughter board. The input signals are timestamped using edge detection and an on-board clock running at 200 MHz. There are multiple input channels since the polarization analyzer used to implement the linear optical Bell-state measurement makes use of multiple detectors, each indicative of a particular polarization state. Thus, the input channel and the timestamp are sufficient metadata to distinguish the state of the detected photons. 
\par
The FPGA writes the generated metadata to an on-board memory region that is also accessible from the ARM processor. The ARM processor uses read/write access to the memory region during execution of user-defined code. This includes, for example, forwarding the recorded metadata to the software-layer. We use the ARM to monitor the local memory buffer for metadata update and program it to respond to request from the network-connected software client. The raw timestamps may be transmitted using either a simple point-to-point UDP broadcast, or they may be transmitted across larger network using the TCP protocol. The implementation reported here uses TCP. 
\par
The transmitted packets include a set of timestamps and channel ids as well as the necessary network overhead, e.g., checksums, packets id, etc. These are received by the software client and processed by the QITKAT program. For the \srccode{SDC Decode} block, this includes correlating different channels with near simultaneous timestamps as a means of identifying a measured Bell state. This task could also be assigned to the middleware, but our chosen API makes this information available to the software layer in order to provide diagnostic monitoring.  
\par
We have not yet implemented a complete build of the TX side. At present, our design suggests suing the ARM-driven FPGA to output control signals that drive the waveplate rotators. Because the ARM executes instruction received from the software-layer via a QITKAT program, it can overwrite the shared memory with the FPGA to modulate the rotator control signals. The more difficult engineering challenge is synchronization of the rotator configuration with the photon pair source. As mentioned above, heralded photon production would provide the necessary timing information, but these are conditional sources that will require the ARM to monitor the availability of a photon pair. Future on demand photon sources will alleviate this requirement.
\section{Super-dense Coding System Simulation Studies}
\label{sec:simstudies}
In the absence of the TX HW, we have tested the super-dense coding system design using numerical simulations of the software and middleware behavior. This includes tracking and storing individual quantum states as well as reproducing the metadata generated by a simulated measurement process. We use these studies to verify the QITKAT implementation of the super-dense coding protocol by checking that an input message can be successfully encoded, decoded, and received.  We use these studies to validate the interaction between the software and middleware, which is based on a client-server model. Both the TX and RX SW interface with a qubit management server (\textit{QM Server}) by sending requests specific to individual qubit transmissions. 
\par
For the RX implementation, we allocated the qubit management server to the FPGA+ARM board. For the numerical studies, we allocate the server to a simulation program running on a separate computer. It is possible to host the client and server on the same computer, but we are explicitly interested in the networking issues that the client-server model must overcome. For each encoded bit pair, the TX SW pushes a two-bit metadata string describing the requested data encoding operation. The server responds to this request with an identification number labeling the prepared qubit. Internally, the simulation tracks only the two-bit metadata and the 32-bit integer label. Similarly, the receiver requests updates from the server about available measurement results. Measurements are simulated assuming an unbiased, complete Bell-state measurement station. This idealization supports our effort to verify the correctness of the software layer and networking between terminals.
\begin{figure}[h]
\begin{center}
\includegraphics[width=.8\textwidth]{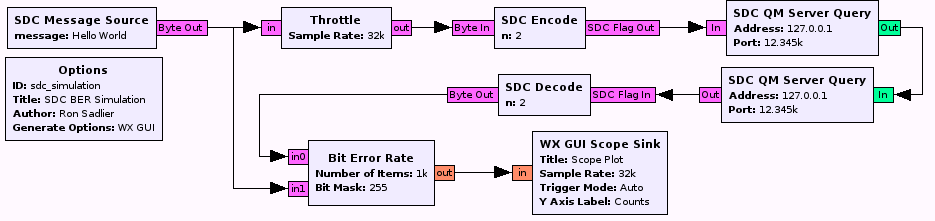}
\caption{The complete QITKAT flow graph for simulating the super-dense coding protocol and verifying the transmission by computing the bit error rate.}
\label{fig:sdcrx}
\end{center}
\end{figure}
\par
\begin{figure}
\centering
\begin{subfigure}[b]{0.45\textwidth}
\includegraphics[width=1\textwidth]{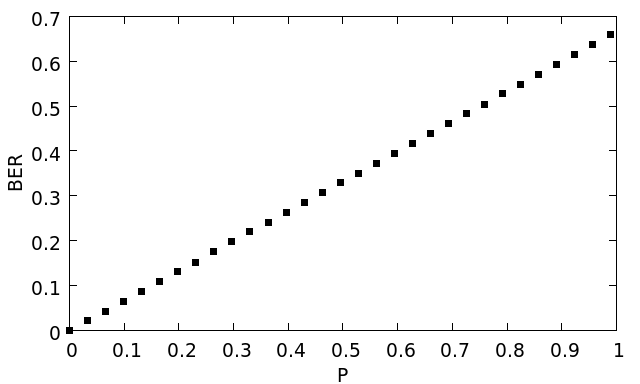}
\end{subfigure}%
~
\begin{subfigure}[b]{0.45\textwidth}
\includegraphics[width=1\textwidth]{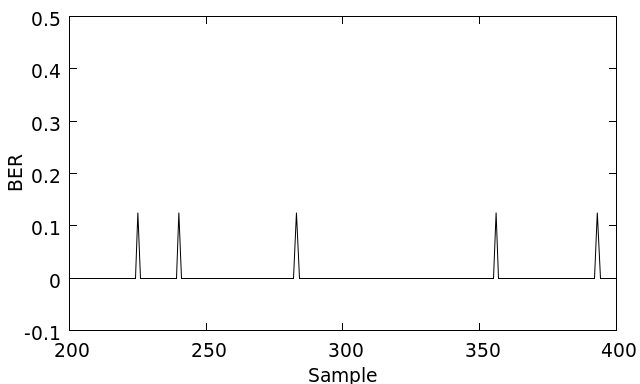}
\end{subfigure}
\caption{(left) Plot of the bit-error rate (BER) observed using numerical simulation of the \srccode{QM Server} and depolarizing channel noise. (right) Plot of the windowed bit error rate over 200 consecutive samples for a fixed value of $p = 0.01$ in the depolarizing noise model.}
\label{fig:ber_results}
\end{figure}

In Fig.~\ref{fig:ber_results}, we present statistical measures of the flow graph from Fig.~\ref{fig:sdcrx} using a numerical simulation of the \srccode{QM Server} block. Our simulation models transmission of the qubit through an isotropic depolarizing noise channel. The depolarizing noise channel transmits the unmodified input state with probability $(1-p)$ and applies each of the Pauli operators $(X, Y, \textrm{ or } Z)$ with probability $p/3$ \cite{Adami1997}. For depolarizing noise, the bit error rate scales linearly in $p$ as $2p/3$, which is precisely the behavior recovered in the first plot of Fig.~\ref{fig:ber_results}. In the second plot of Fig.~\ref{fig:ber_results}, we show a snapshot of the transmission by recording the windowed bit error rate over a range of 200 consecutive samples for a fixed noise parameter $p=0.01$. The bit error rate is nominally zero but spikes occasionally when an incorrectly encoded state is received.
\section{Conclusions}
We have extended the paradigm of software-defined communication to the context of a point-to-point QC system. We defined a layered model for the transmitter and receiver that separates each QC terminal into hardware, software, and middleware concerns. Our  design methodology emphasized the role of middleware for abstracting the high-level, software control language and managing the low-level hardware operations. We gave a detailed description of how each layer operates as well as a concrete implementation based on parts commonly found in existing QC designs. 
\par
We have used the SDQC framework to design a super-dense coding system. Our approach includes an extension of GNU Radio for the software layer and an FPGA+ARM-based solution for the middleware layer. In the absence of experimental hardware, we have emulated the middleware behavior at each terminal using numerical simulation to model the transmission and measurement of  quantum states. We tested our implementation for correctness as well as behavior with respect to the dimensionality of the transmitted quantum state. These results have been used to validate the software layer and provide insights into the classical overhead associated with implementing the protocol.
\par
The motivation for SDQC is to establish design methodologies that enable rapid prototyping of experimental systems. The availability of a reconfigurable software layer supports testing a robust family of communication protocols at significantly less expense in terms of development time. We are currently applying to the case of dense coding using error corrected transmissions \cite{HumbleSadlier2014}.  Because QC is a relatively young field with a large design space, the ability to explore design parameters rapidly using prototype systems supports the testing of new theories and the assessment of existing communication strategies. The versatility of SDQC testbeds is useful for exploring new regimes in communication.
\par
Our discussion of QC systems has been limited to end-user terminals. We have not discussed the implementation of the classical or quantum networks that connect users, apart from assuming that these network exist and that they have well-defined interfaces. Recently, van Meter and Touch have discussed the design decisions underlying quantum networks and inter-networks.\cite{vanMeterTouch2013} They provided insight into the protocols and stacks needed to support the type of interactions not included here. We anticipate it is possible to extend our present ideas to similar concerns, for example to software-defined quantum networking.
\section*{ACKNOWLEDGMENTS}   
T.~S.~H.~thanks Henry Humble for comments regarding the SDQC design in Sec.~\ref{sec:frame}, Toby Flynn and Laura Ann Anderson for help evaluating the Zynq middleware, and Ali Ismail for initial efforts in defining the middleware interface.  R. J. S. thanks the Department of Energy Science Undergraduate Laboratory Internships (SULI) program for support. This work was supported by the Defense Threat Reduction Agency. This manuscript has been authored by a contractor of the U.S. Government under Contract No.~DE-AC05-00OR22725.

%


\vspace{2ex}\noindent{\bf Travis Humble} is co-director of the Quantum Computing Institute at Oak Ridge National Laboratory and an assistant professor at the University of Tennessee. He received his PhD in chemistry from the University of Oregon in 2005. He is the author of more than 50 journal papers. His current research interests include quantum information, communication, and computing. He is a member of SPIE.

\vspace{2ex}\noindent{\bf Ronald Sadlier} is a junior at the University of Rhode Island. He is currently majoring in physics and math with an expected graduation data of 2014. His current research interests include quantum information, communication, and computing.
\vspace{1ex}

\listoffigures
\listoftables

\end{spacing}
\end{document}